\documentclass{ws-jai}
\usepackage[flushleft]{threeparttable}
\usepackage{subcaption}
\usepackage[ruled,vlined]{algorithm2e}
\usepackage{graphicx}

\begin{document}

\catchline{}{}{}{}{} 

\markboth{Alessio Magro}{Digitizing MEXART - System Overview and Verification}

\title{Digitizing MEXART - System Overview and Verification}

\author{A. Magro$^{1,*}$,
J. Borg$^{1}$,
R. Chiello$^{2}$,
D. Cutajar$^{1}$,
K. Zarb-Adami$^{1,2}$, 
J. A. Gonzalez-Esparza$^{3}$,
E. Andrade$^{3}$, \\
E. Aguilar-Rodriguez$^{3}$,
J. C. Mejia-Ambriz$^{3}$ and
P. Villanueva$^{3}$}

\address{
$^{1}$Institute of Space Sciences and Astronomy (ISSA), University of Malta, Msida, MSD 2080, Malta, fauthor@university.com\\
$^{2}$University of Oxford, Denys Wilkinson Building, Oxford, OX1 3RH, United Kingdom \\
$^{3}$LANCE, Instituto de Geof\'{i}sica, Unidad Michoac\'{a}n, Universidad Nacional Aut\'{o}noma de M\'{e}xico, Morelia, Michoac\'{a}n, M\'{e}xico. CP 58190
}

\maketitle

\corres{$^{*}$Corresponding author.}

\begin{abstract}

The Mexican Array Radio Telescope (MEXART), located in the state of Michoacan in Mexico, has been operating in an analog fashion, utilizing a Butler Matrix to generate fixed beams on the sky, since its inception. Calibrating this instrument has proved difficult, leading to loss in sensitivity. It was also a rigid setup, requiring manual intervention and tuning for different observation requirements. The RF system has now been replaced with a digital one. This digital backend is a hybrid system utilizing both FPGA-based technology and GPU acceleration, and is capable of automatically calibrating the different rows of the array, as well as generating a configurable number of frequency-domain synthesized beams to towards selected locations on the sky. A monitoring and control system, together with a full-featured web-based front-end, has also been developed, greatly simplifying the interaction with the instrument. This paper presents the design, implementation and deployment of the new digital backend, including preliminary analysis of system performance and stability.

\end{abstract}

\keywords{instrumentation: interferometers; instrumentation: radio astronomy; techniques: digital signal processing; methods: observational}




\section{Introduction}
\label{introduction}

The Mexican Array Radio Telescope (MEXART) is a transit radio interferometer consisting of a regular grid of 64x64 full wavelength dipoles located in Michoacán, Mexico. The primary scientific purpose of this instrument is to measure extragalactic radio sources around the Sun to track large-scale solar wind disturbances propagating between the Sun and the Earth \citep{gonzales:mexart}, working at an operating frequency of 139.65 MHz. The array has 64 East-West (E-W) rows, each of which contains 64 equally spaced dipoles, partitioned into four sections. 

MEXART was originally an analog-based telescope, using a 16x16 Butler Matrix beamformer to generate 16 fixed latitudinal beams at different declinations. The dipoles of each row are combined through several combination stages, resulting in one analog output per row. Since the Butler Matrix only has 16 inputs, adjacent rows had to be combined together to use half, or the full, array. This resulted in poor beam directivity, particularly due to the fact that the combined signals could not be easily gain and phase calibrated. The calibration process was performed manually through analog equipment, and required a strong astronomical source to transit the telescope in order to get a good enough signal-to-noise ratio (SNR) to determine the appropriate correction factor for a specific array section.  This has prevented the instrument from achieving the required sensitivity to detect the number of sources necessary for solar wind study. 

An upgrade effort was started in 2019, where the Butler Matrix was replaced by a digital and software backend. The first phase of the upgrade effort is documented in \citet{magro:mexart}, where the preliminary architecture is described, including the Radio Frequency (RF) conditioning required to connect the combined row signals to the digital backend. The work presented here details the real-time processing pipeline and provides preliminary verification of the upgraded system through astronomical observations. 

\begin{figure}
\begin{centering}
  \includegraphics[width=260pt]{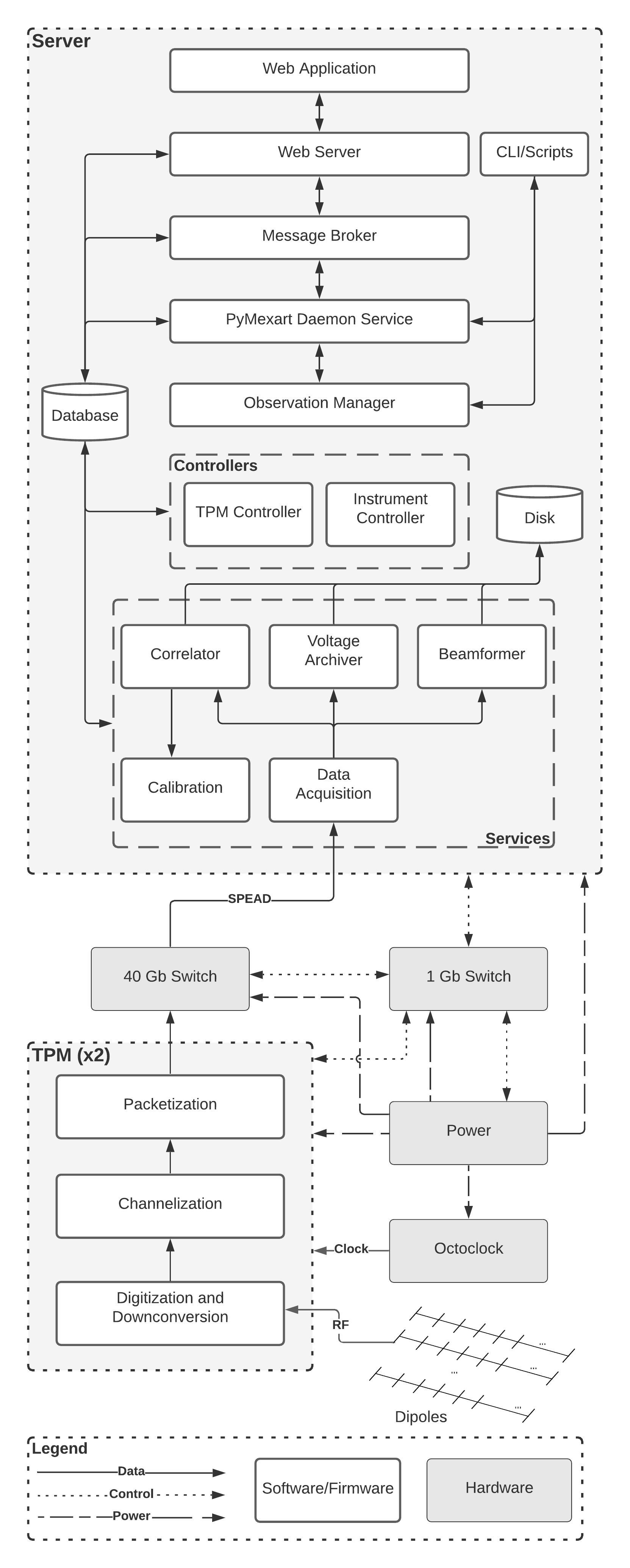}
    \caption{Digital system overview, including FPGA, GPU and general software components.}
    \label{fig:system_overview}
    \end{centering}
\end{figure}

Figure \ref{fig:system_overview} shows an overview of the deployed system. The signal from each of the 64 rows, after undergoing RF amplification and filtering (see \citet{magro:mexart}, Section 2) are input into the digital backend, which consists of 2 Tile Processing Modules (TPMs) \citep{naldi:tpm}. Each RF signal is digitized, down-converted, channelized and transmitted to the processing server over a 40 Gb/s Ethernet link using a custom SPEAD (\citet{spead}) packet structure adapted from \citet{magro:daq}. Each TPM has two 40 Gb/s network interfaces (one for each field-programmable gate array (FPGA)), both of which are connected to the 40 Gb switch. Since the cumulative data rate from both TPMs is relatively low, totaling approximately 13 Gb/s, only a single connection is required between the switch and the compute server, with an additional link for redundancy.

The server hosts an NVidia Tesla P40 Graphics Processing Unit (GPU) which, during normal operation, generates a number of synthesized beams within the array's primary Field of View (FoV) in real-time using a custom GPU-based frequency-domain coherent beamformer, described in Section \ref{beamformer}, with the resulting beams saved to disk in HDF5\footnote{\url{https://www.hdfgroup.org/solutions/hdf5/}} format for offline processing. To compensate for instrumental delays which decohere the signals, a calibration observation is routinely performed using a custom GPU-based correlator that calculates the correlation matrix for the array. This is then used by the calibration routine to generate per-antenna gain and phase solutions and used subsequently during beamforming observations. Apart from the real-time processing software, the server also hosts the software infrastructure for observation management, instrument monitoring and control, and the web-based front-end. 

The system also consists of support hardware, including two separate networks: 1 Gb control network with two switches for communicating with all the hardware in the rack, and a 40 Gb data network with one switch, connecting the output of the TPMs to the server. Two power distribution units (PDU) provide power to all the equipment. The PDUs are connected to the 1 Gb network so that individual devices can be powered up or down remotely. A fan-based cooling system surrounds the TPMs, whilst an OctoClock G\footnote{\url{https://www.ettus.com/all-products/octoclock-g/}} provides pulse-per-second (PPS) and 10 MHz signals to the TPMs. 

The rest of this paper describes the new system in greater detail, starting with the Digital Backend in Section \ref{digital}, followed by the processing and control system in Section \ref{software}, which includes the beamformer, correlator, calibration and the front-end. In Section \ref{system_verification} we provide a preliminary analysis of the performance and stability of the new instrument, with Section \ref{conclusion} concluding this paper.

\section{Digital backend}
\label{digital}

\begin{figure*}
  \includegraphics[width=\textwidth]{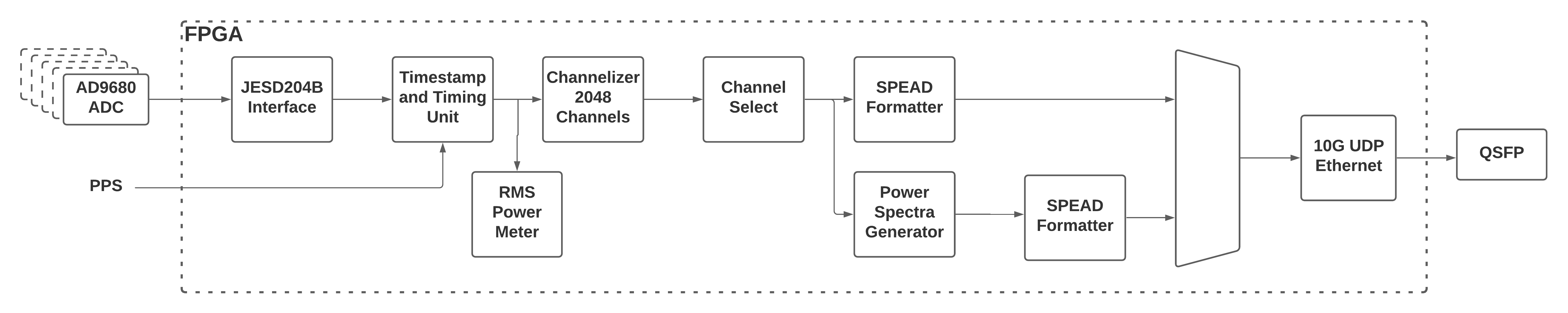}
    \caption{FPGA Firmware block diagram. Each FPGA processes 16 digitized RF signals producing channelized data divided into 512 frequency channels with a bandwidth of 12.5 MHz. Two TPM boards, hosting two FPGAs each, operate in parallel in order to process 64 RF signals. Channelized data and integrated spectra are fragmented into SPEAD packets and transmitted to the processing server. Adapted from \citet{magro:mexart}.}
    \label{fig:firmware}
\end{figure*}

The 64 RF-conditioned signals are transported to the analogue inputs of two TPMs which run the custom firmware developed for MEXART. Figure \ref{fig:firmware} shows the FPGA firmware block diagram. The two FPGAs on the digital board are connected to eight dual channel AD9680 Analog-to-Digital Converters (ADCs), which operate at 800 MSPS and are capable of processing the digitized signals using integrated Digital Down Converters (DDCs). MEXART's original processable bandwidth was around 2 MHz, centered at 139.65 MHz. The desired bandwidth of the digital system was approximately 10 MHz. To achieve this, the central bandwidth is downconverted using the DDCs, resulting in a sampling rate of 100 MSPS, corresponding to an observable bandwidth of 50 MHz. The digitized and downconverted signals are then transferred to the FPGAs and grouped into frames of 4096 samples. Each frame is timestamped by taking into account the Epoch time and the rising edge of the Pulse-per-second (PPS) (the Epoch time is set during system initialization and synchronization).

The timestamped frames are then processed by a Polyphase Filter Bank Channelizer which produces 2048 frequency channels. Since the desired observable band is 12.5 MHz, only 512 channel are required. The Channel Select module can be instructed to select 512 contiguous frequency channels from the generated 2048 for further processing. The final channel bandwidth is therefore 24.4 kHz. After channelization, data are processed by the SPEAD formatter and forwarded to the 10G UDP Ethernet core in order to be transmitted over the network to the processing server. 

Several diagnostic functions have also been implemented in the FPGA firmware. A snapshot of ADC data from all antennas can be transmitted to the server for inspection. A Root Mean Square (RMS) power meter calculates real-time RMS power of the time domain DC data for all input signals, allowing for antenna power monitoring and broadband RFI detection. Additionally, the power spectrum of each antenna can be calculated by integrating the output of the channelizer for a configurable time period. The generated power spectra are then transmitted over the network at a low data rate to the processing server, where they are stored. These spectra are displayed in the monitoring page of the front-end where users can visually determine the health of the instrument.

The TPMs and control software communicate by using the Python FPGA Board Interfacing Layer (PyFABIL), which was developed for the Square Kilometer Array (SKA) \citet{magro:pyfabil}. Memory-mapped registers and memory areas in the FPGA and other on-board devices are exposed through an automatically generated XML file during the compilation stage. The host can read this file (which is memory mapped by the firmware at a specific address) and create an internal representation of the board and firmware, so that register names can be automatically converted to the appropriate addresses. The library includes a plugin system to implement firmware-specific functionality. A few plugins were added to PyFABIL for the MEXART design, including the interface to the DDCs and Channel Select, while the rest were reused from the SKA prototype implementation. In the MEXART setup, all monitoring and control communication is performed over the 1 Gb network.

\section{Processing and control system}
\label{software}

The processing server hosts the MEXART software suite, the architecture of which is shown in Figure \ref{fig:system_overview}. All SPEAD packets are transmitted to a single 40 Gb interface, whilst all control data flows across the 1 Gb network. When an observation is running, an instance of each required Service is created. The Data Acquisition component, which reads the incoming SPEAD packet stream, is required for all Services that process the data in real-time and generate HDF5 files containing the output products. Observations are managed by the Observation Manager, which includes the scheduler. The Daemon Service executes as a system service and is responsible for setting up the required processing pipeline for each observation, cleaning up after an observation finishes, and coordinates the monitoring and control activities which are performed though hardware-specific controllers. The TPM controller manages the TPMs, including programming and initialization, configuration and  periodic acqusition of antenna spectra and RMS power, which are stored in the database. The database also stores the observation schedule and history, list of sources and calibration coefficients. A user can interact with this system either via the web application, which is decoupled from the rest of the architecture by a REDIS\footnote{\url{https://redis.io/}}-based message broker, or through several command-line scripts. The following subsections describes several components of the MEXART software suite in detail.

\subsection{Processing pipeline}

\begin{figure*}
  \includegraphics[width=\textwidth]{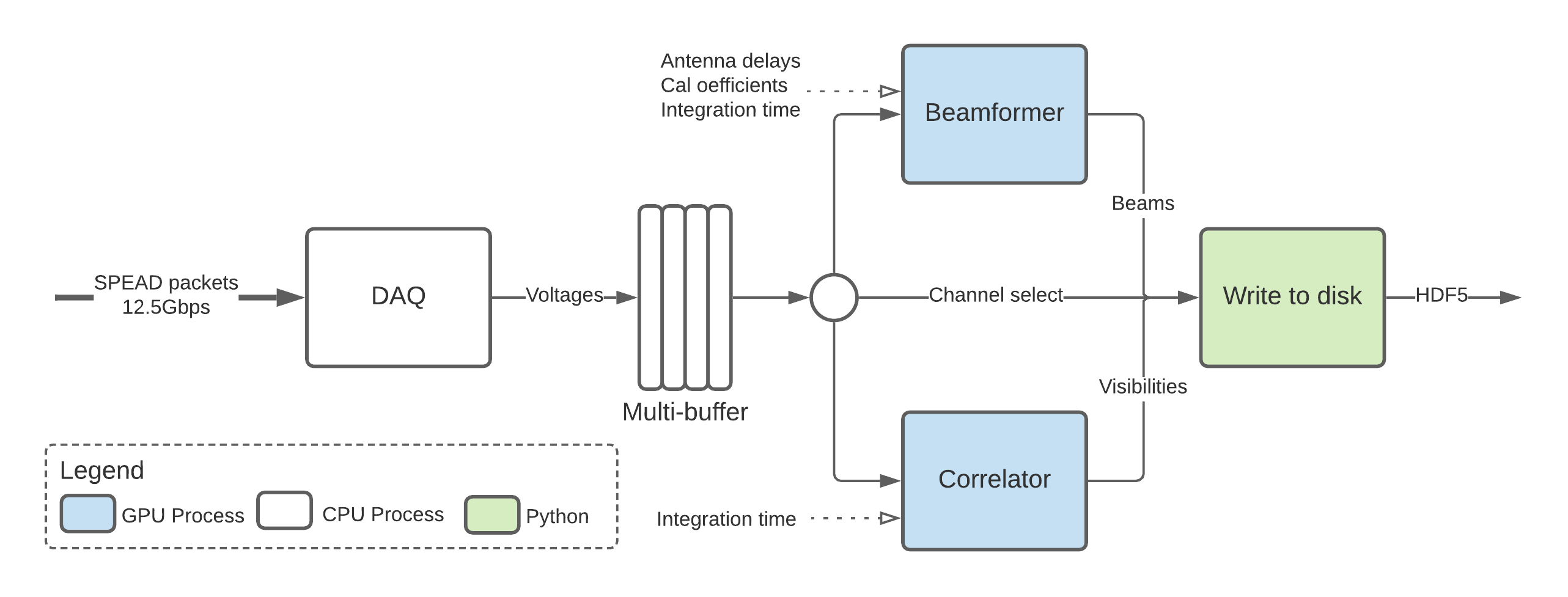}
    \caption{Data processing pipeline. Incoming SPEAD packets are received, filtered and extracted by the DAQ process, which forward the packet content to a buffering system. Consumers of this data, namely the beamformer, correlator and the voltage archiver, process this data to generate the required output products, which are saved to disk. See text for more details.}
    \label{fig:processing}
\end{figure*}

Performing an observation involves creating the required processing pipeline to process the incoming channelised voltages in real-time. A pipeline is composed of three primary components: the data acquisition component, which handles packet reception and buffering, the processing component, which either beamforms or correlates the data, and the storage component, which saves the generated output to disk. The data acquisition component, which will henceforth be referred to as the DAQ, is the core DAQ library described in \citet{magro:daq}. The beamformer, correlator and voltage archiver are implemented as packet consumers which register themselves with the DAQ and wait for filtered packets to process, a general structure of which is show in Figure \ref{fig:processing}. These pipelines are wrapped in a Python layer that can create and configures the pipelines and receives the generated output, which is saved as HDF5 files to disk. The files contain two datasets sets, one having the actual processed datasets and associated Coordinated Universal Time (UTC) timestamps for each data point, and another containing the observation metadata. For long observations, several self-contained HDF5 files are generated, the size of which is configurable.

\subsubsection{Data acquisition}
\label{daq}

An adapted version of the high-performance data acquisition library described in \citet{magro:daq} is used to capture the high-speed SPEAD stream from the TPMs for real-time processing. This library makes use of a Linux kernel feature which allows sets of raw Ethernet frames to be buffered in a ring buffer residing in kernel memory that is memory mapped to user memory, thus avoiding any system calls that would severely degrade the performance of the data acquisition system. It allows several packet consumers to be registered with the networking thread(s), creating a ring-buffer between the two. Each consumer provides a packet filter which the network thread uses to determine which packets should be pushed to the ring buffer. The DAQ has been stress tested to work up to 80 Gb/s with multiple receiving threads. The data rate in MEXART is approximately 12.5 Gb/s which can he handled by a single receiving thread. However, in order to reduce packet loss, this thread needs to be mapped to a CPU core which is close to the PCIe interface to which the 40 Gb network card is attached. Additionally, consumers attached to the network receiver are placed next to this core (at least on the same CPU) to avoid NUMA memory latency, and reduce cache synchronization and invalidation overhead. The beamformer, correlator and voltage archiver are implemented as DAQ consumers, and are described below.

\subsubsection{Beamformer}
\label{beamformer}

The beamformer can be regarded as the core processing element of the MEXART software suite. Due to the high data-rate and requirement for generating several synthesized beams, the beamformer is implemented as a CUDA \citep{cuda} kernel. The beamformer consumer is implemented as two separate threads: one responsible for buffering and another responsible for copying data in and out of the GPU and running the beamforming kernel. The buffering thread generates a multi-buffer (by default containing 4 buffers) whose size depends on the configuration of the beamformer, such as the integration time and number of integrations per call. When SPEAD packets are available on the ring buffer the buffering thread processes the packets and places the contents at the appropriate offset within the multi-buffer. Using a multi-buffer allows for handling out-of-order packets at buffer boundaries. When a buffer is flagged as ready the beamforming thread is notified and the contents are copied to GPU memory, after which the GPU kernel is run. Once the beams are generated they are copied out of the GPU and forwarded to the Python interface layer via the callback mechanism. 

Beamforming can be parallelized across frequency and pointing, so a 3D CUDA grid is used, with dimensions $(N_{\texttt{block}}, N_{\texttt{freqs}}, N_{\texttt{beams}})$, where $N_{\texttt{blocks}}$ is the number of blocks to process, $N_{\texttt{freqs}}$ is the number of frequency channels and $N_{\texttt{beams}}$ is the number of synthesized beams to generate. $N_{\texttt{blocks}}$ is simply $N_{\texttt{time}}$, which is the total number of samples to process, divided by $N_{\texttt{threads}}$, which is the number of threads in each CUDA thread block (one dimensional thread blocks are used). Therefore, each CUDA thread is responsible for generating one synthesized beam, for one frequency channel, for one time sample. Integration is then performed across threads, and possibly across thread blocks, as described shortly.

\begin{algorithm}
\SetAlgoLined
\KwResult{Beamformed voltages}
 Cooperative load of antenna voltages\;
 Cooperative load of calibration coefficients\;
 Cooperative calculation of frequency-dependent delay\;
 Cooperative application of delay and calibration coefficients\;
 Synchronize\; 
 \BlankLine
 \For{All antennas}{
    Unpack antenna voltage\;
    Apply pointing and calibration coefficient to antenna\;
    Add to beamformed value\;
 } 
 \BlankLine
 \uIf{integrate >= warpSize} {
    Warp-level shuffle to combine beamformed values\;
    Atomic-add to global memory by 1 thread\;
 }\Else {
    Atomic-add to global memory by each thread
 }
\caption{Beamforming kernel}
\label{algo:beamformer}
\end{algorithm}

A high-level algorithmic description of the GPU-based beamformer is provided in Algorithm \ref{algo:beamformer}. The first operation is to  compute the phase delays for pointing the synthesized beams. A set of antenna phase shifts is provided to the kernel, which then calculates the frequency-dependent phase delays for the synthesized beams which need to be generated. Since MEXART is a transient instrument, the antenna phase shifts do not need to be updated during an observation, however, the frequency-dependent delays are computed within the kernel to reduce the number of global memory accesses required (it's faster to generate these delays every time rather than loading them from memory). The phase shifts are stored in constant memory. All threads in a thread block participate to generate the phase delays for all antennas, one frequency channel and one beam, across $N_{\texttt{threads}}$.

Within the same operation, the gain and phase coefficients are applied. These coefficients are copied to GPU global memory during observation initialisation. When a phase delayed antenna sample is computed, it is multiplied by the corresponding calibration coefficient and stored to a shared memory array.

Once the antennas are pointed and calibrated they are coherently combined. $N_{\texttt{threads}}$ summations are performed concurrently in a thread block, with the partial beams stored in a shared memory  array. This array has dimensions $(N_{\texttt{nants}}, N_{\texttt{threads}} + 1)$, where $N_{\texttt{ants}}$ is the number of antennas. A one is added to $N_{\texttt{threads}}$ to reduce the number of bank conflicts when accessing the shared memory area. Once the full beams are generated, the complex components are squared and combined. 

The final step is time integration. If the number of samples to integrate is larger than the warp size (generally 32), then CUDA warp-level primitives are used to combine the computed beams in a tree-reduction manner, with the final result saved atomically to global memory. If the number of samples is less than the warp size, then each thread performs an atomic addition in global memory.

When the beamforming kernel finishes execution, the generated beams are copied to CPU memory and the consumer callback is called. The callback is implemented within the Python wrapper. In this callback the values are written to an HDF5 file, which contains both the beam data and observation metadata that includes the pointings in altitude and azimuth, observed source (if set), timestamps for each time sample, frequency range and other observation parameters.

\begin{figure}
\begin{centering}
   \includegraphics[width=320pt]{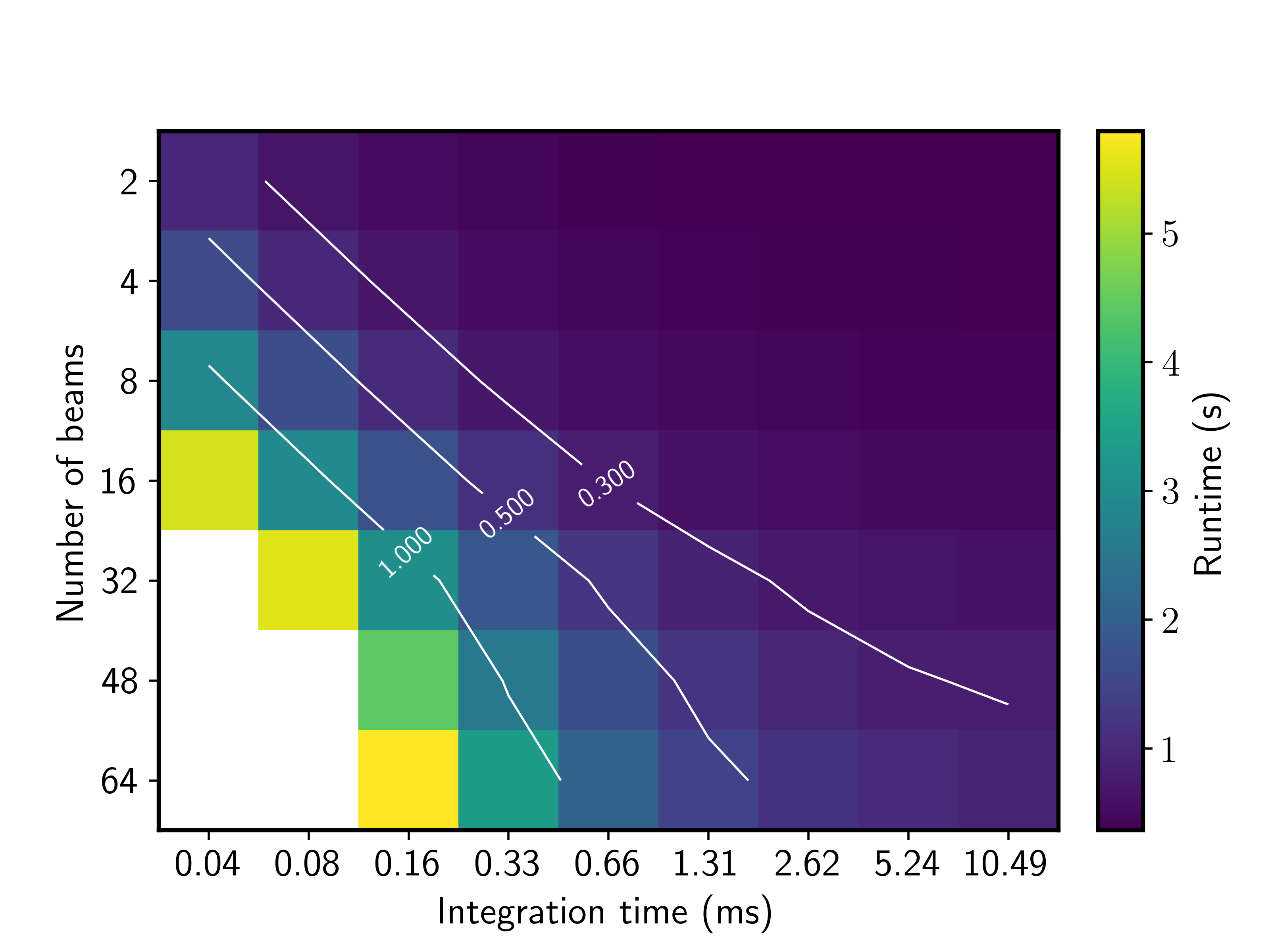}
    \caption{Beamforming kernel benchmark for combinations of number of beams and integration time. The contour lines represent factor of real-time, where a value $\leq{1}$ signifies that the kernel can process the data in realtime. The buffer length (number of samples) was fixed to 65536, or $\sim$2.68 s, and the benchmarking GPU is an NVidia V100 SXM2. The timings assume a disk write speed of 450 MB/s.}
    \label{fig:beamformer_benchmark}
\end{centering}
\end{figure}

The beamformer can currently be configured to generate up to 64 synthesized beams, with an integration time range of 0.1 ms to 2 s. The number of beams which can be generated in real-time depends on the configured integration time and disk write speed. The lower the integration time the more beamformed samples need to be copied out of GPU memory and saved to disk, and this increases linearly with the number of beams. Figure \ref{fig:beamformer_benchmark} show the runtime, in seconds, of the beamformer for a combination of number of beams and integration time. The buffer length for the benchmarks was fixed to 65536, equivalent to $\sim$2.68 s of observation time. The superimposed contour lines represent fractions of real-time, such that a value of 1.0 means that the kernel can process the combinations at exactly real-time, while a value of 0.5 means that the kernel is twice as fast. These benchmarks assume a disk-write speed of about 450 MB/s, which is the benchmarked write speed on the MEXART server. The white pixels in the lower left hand corner of the plot represent combinations which do not fit in GPU memory for the selected buffer length. These benchmarks were performed on an NVidia V100 SXM2 GPU.

\subsubsection{Correlator}
\label{correlator}

Calibration observations require a correlator to generate cross-correlation matrices, or visibilities, which are then used by the calibration algorithm to generate the calibration coefficients used to calibrate the synthesized beams. Since the data rate is too high to write the data directly to disk, the antenna signals have to be correlated in real time, and a custom GPU-based correlator was developed for this task. The same concept as the GPU beamformer is used, with a buffering thread and a GPU thread. Originally the intention was to use xGPU (\citet{clark:xgpu}), however, some issues with single polarization mode arose, and since the required compute rate could be handled by a naive correlator implementation on the GPU, an in-house implementation was used.

The correlator generates an integrated correlation matrix with a configurable integration time. To keep the implementation of the correlator simple, only the number of samples needed to generate one integrated sample are copied into the GPU, such that the kernel generates a single correlation matrix (per frequency channel) in a call. Therefore, the execution only needs to be parallelised across two dimensions, baseline and frequency channel. Each CUDA thread is responsible for generating the integrated cross correlation for one baseline and one frequency channel, such that one 1D thread block can handle a subset of baselines for one channel. A 2D CUDA grid is used, with baselines subsets (total number of baselines divided by number of baselines processed in a thread block) along the first dimension and frequency channels along the second dimension. 

As with the beamforming kernel, when the correlator finishes execution, the visibilities are copied to CPU memory and the consumer callback is called. This callback performs the same operations, albeit with a different data organization in the HDF5 file.

\subsubsection{Voltage archiver}
\label{voltage_archiver}

The final consumer is the voltage archiver, which can continuously write a single frequency channel to disk. This mode is used primarily for testing and validation, as offline versions of the beamformer and correlator can be tested on offline data. Generally, a validation observation is performed by observing a bright astronomical source and storing the central frequency channel to disk, following the same mechanism as the beamformer and correlator.

\subsection{Calibration}
\label{calibration}

Radio interferometers, or antenna arrays in general, have instrumental errors which need to be accounted for in order to get maximum SNR when combining the antenna signals to generate beams. These delays are antenna-specific and unpredictable, and can vary in an erratic manner with time. Sources of instrumental errors include different cable lengths, electronic circuitry, analogue filters and amplifiers, and environmental factors such as temperature and humidity. To compensate for instrumental errors, calibration observations need to be performed routinely using strong calibrator sources (point sources with high flux, such as Cassiopeia A, Cygnus A, Taurus A and 3C123), allowing for the calculation of deviations from the expected observer visibilities. For these observations, the MEXART system is used in correlation mode, where the antenna voltages are correlated in real-time as the calibrator source transits the meridian. These observations typically last five to ten minutes. The generated correlation matrix is then input to the calibration algorithm, which generates the calibration coefficients that are used by the beamformer to form the coherent synthesized beams.

The calibration routine implemented in MEXART is StEFCal \citep{salvini:stefcal}, which is a minimization algorithm aimed at reducing differences between model visibilities and uncalibrated visibilities, to an acceptable tolerance level, by computing best-fit-per-element coefficients. In the current implementation, the model visibilities simply assume that all antennas will have zero phase difference at the time of transit, after the geometrical components are removed from the observed visibilities, such that any residual phase is due to instrumental errors. Therefore, there is an assumption that perfect dipoles are being used, with negligible mutual coupling effects between different dipoles, and thus no element or average element patterns are required to generate the model visibilities. 

The calibration procedure starts off by selecting three visibility sets from the correlation observation, one at the exact time of transit, one a minute prior and one a minute after the transit time. The same calibration procedure is applied to the three sets, and the one resulting in the best calibration coefficients is then used. This provides some resilience to Radio Frequency Interference (RFI). The best calibration coefficients are the ones having the least gain variance. For each visibility set, StEFCal is run on each of the frequency channels, resulting in $N_{\texttt{nfreqs}} \times N_{\texttt{nants}}$ coefficients. The coefficients of any unwanted antennas (either set by the observer, or automatically ignored by checking the phase values) are set to zero. After selecting the best coefficients, the geometrical component is removed and the resulting values are saved in the database and on disk.

\subsection{Front-end}
\label{system_control}

The MEXART radio telescope can be monitored and controlled through a web-based front-end. The operator can monitor the spectra received from the instrument's two TPM modules. In addition, one can schedule both drift (beamformer) and calibration (correlator) observations through this cohesive interface. Communication between the front-end and the data processing backend is achieved through a  communication layer that uses a REDIS database as a message broker between the two. Control messages are exchanged between the backend and front-end processes in a bi-directional fashion. These messages are encoded as JSON objects and are published to the corresponding channels on the message broker. For instance, an event message is consumed by a subscriber, such as a web application, listening for messages on the corresponding channel. This event message, which could range from a fatal error to an 'observation scheduled' event,  is in turn visualised as a notification on the web portal. This approach makes the front-end completely decoupled from the underlying data processing system. This facilitates the re-use and deployment of the software on other radio instruments. A similar approach has already seen extensive use in other telescope installations, such as the BIRALES UHF radar (\citet{denis:pybirales}).

\section{System Verification}
\label{system_verification}

\begin{figure*}
     \centering
     \begin{subfigure}[b]{0.45\textwidth}
         \centering
         \includegraphics[width=\textwidth]{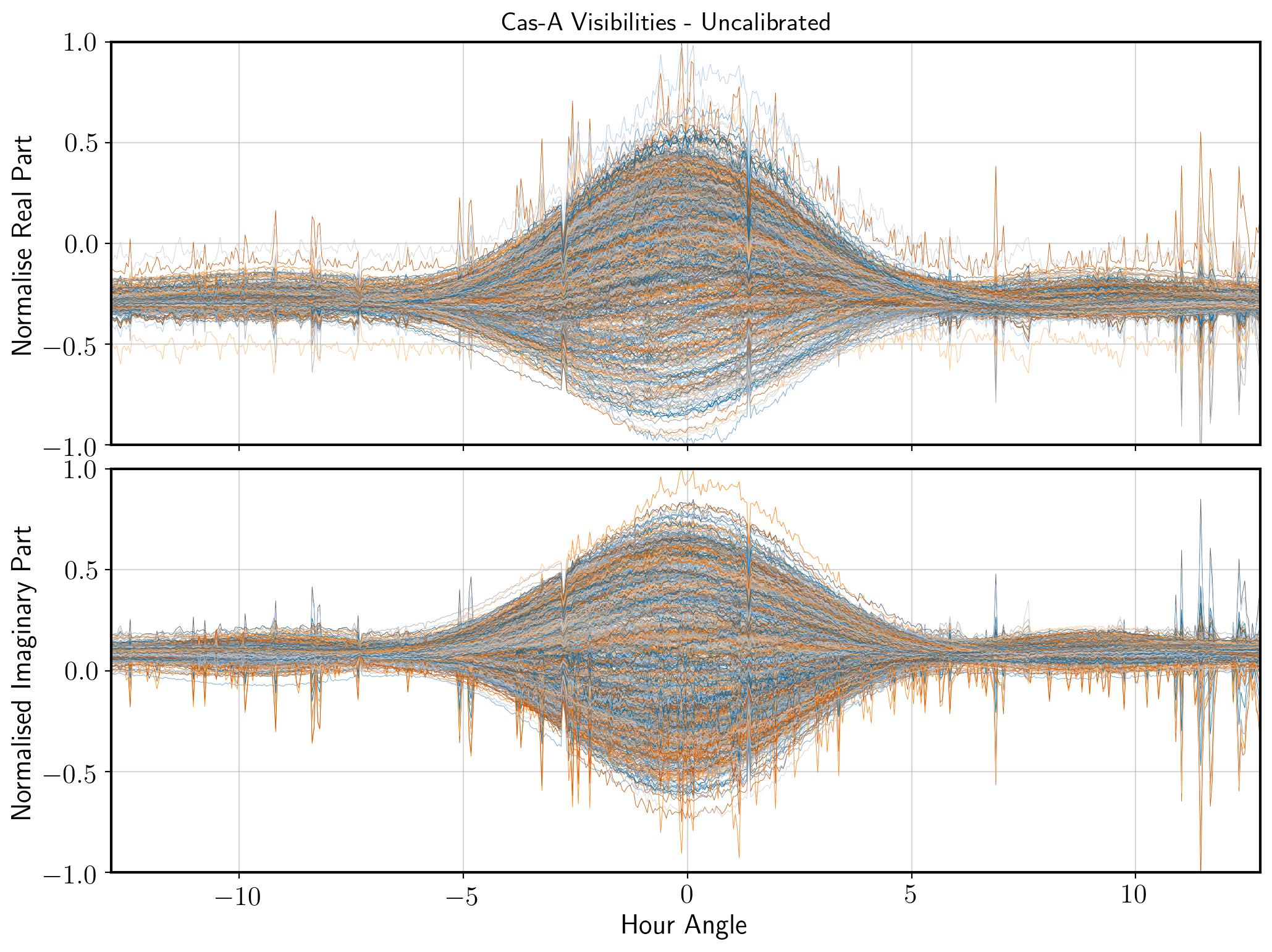}
         \caption{Cassiopeia A observation with geometric coefficients.}
         \label{fig:casa_uncalibrated}
     \end{subfigure}
     \hfill
     \begin{subfigure}[b]{0.45\textwidth}
         \centering
         \includegraphics[width=\textwidth]{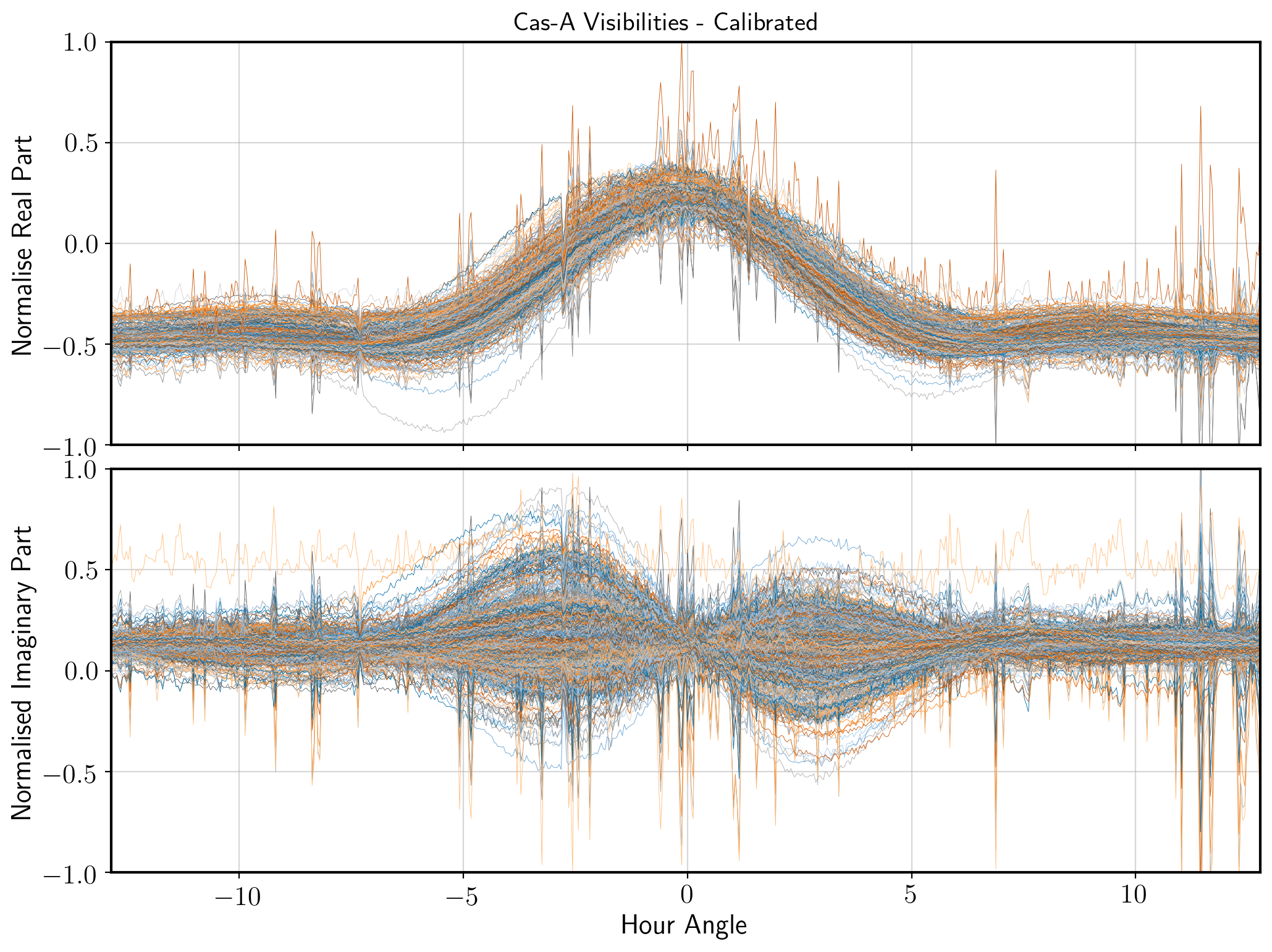}
         \caption{Cassiopeia A observation with calibration and geometric coefficients.}
         \label{fig:casa_calibrated}
     \end{subfigure}
        \caption{A 30 minute calibration observation of Cassiopeia A, showing the real and imaginary components of all visibilities excluding auto-correlations, with (a) having geometric coefficients such that the visibilities are phased towards the source, and (b) having both geometric and calibration coefficients applied.}
        \label{fig:casa_observation}
\end{figure*}

Incremental deployment of the digital backend started in November 2019. The equipment cabinet was set up and one TPM was installed. Half of the array, the 32 rows corresponding to the southern part of the telescope, was connected to the TPM and a series of tests on the RF quality of the signals were conducted. Astronomical observations were subsequently performed, and by the middle of 2020 the second TPM was installed for the northern part, with the full array connected, at which point system synchronization and timing was verified. For the rest of 2020 and beginning of 2021 the digital system has undergone continuous testing and verification and, finally, astronomical observations (\citet{americo:mexart}). 

Figure \ref{fig:casa_observation} shows the effect of calibrating an observation of Cassiopeia A. The center frequency, a 24.4 KHz bandwidth channel, was recorded to disk using the Voltage archiver for 30 minutes around the source transit time on 13\textsuperscript{th} January 2021 at 22:36 UTC. The correlator was used in offline mode to generate the correlation matrix, or visibilities, which were then input to the calibration module to generate the coefficients.  Figure \ref{fig:casa_uncalibrated} show the visibilities with only geometric components, such that they are phased towards the location of Cassiopeia A, whilst for Figure \ref{fig:casa_calibrated} the generated calibration coefficients were also applied. The fringes show the main peak at around transit ($\omega = 0^{\circ}$). Each colour corresponds to one of the 2080 independent baselines. The spikes in the fringes represent RFI events, which this observation was not compensated for. The imaginary part for all the baselines approaches zero at transit time, signifying that there is minimal phase delay between the antennas towards the phased direction (the geometric pointing). The calibrated fringes do not align perfectly, there is significant scatter around the zero value of the imaginary part and maximum value (excluding the RFI spike) of the real part. This could be due to the fact that each antenna is in reality 64 dipoles going through several analog combination steps which do not compensate instrumental phase offsets within the row.

\begin{figure*}
  \centering
  \begin{subfigure}[t]{0.45\textwidth}
    \includegraphics[width=\textwidth]{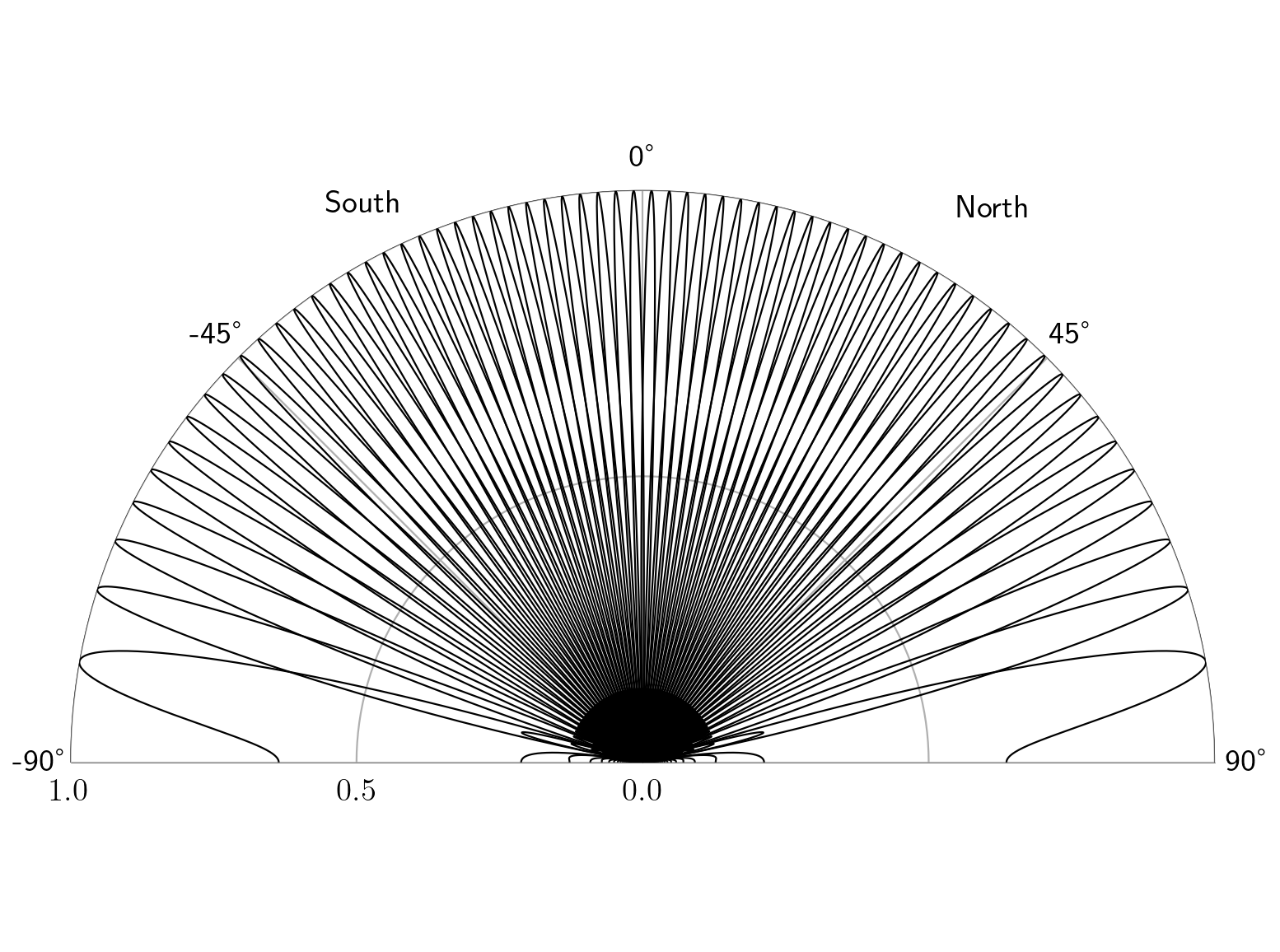}
      \caption{Pointing configuration of the synthesized beams, from beam 1 to 64, used for verification. Adapted from \citet{americo:mexart}.}
      \label{fig:beam_pattern}
  \end{subfigure}
  \begin{subfigure}[t]{0.45\textwidth}
    \includegraphics[width=\textwidth]{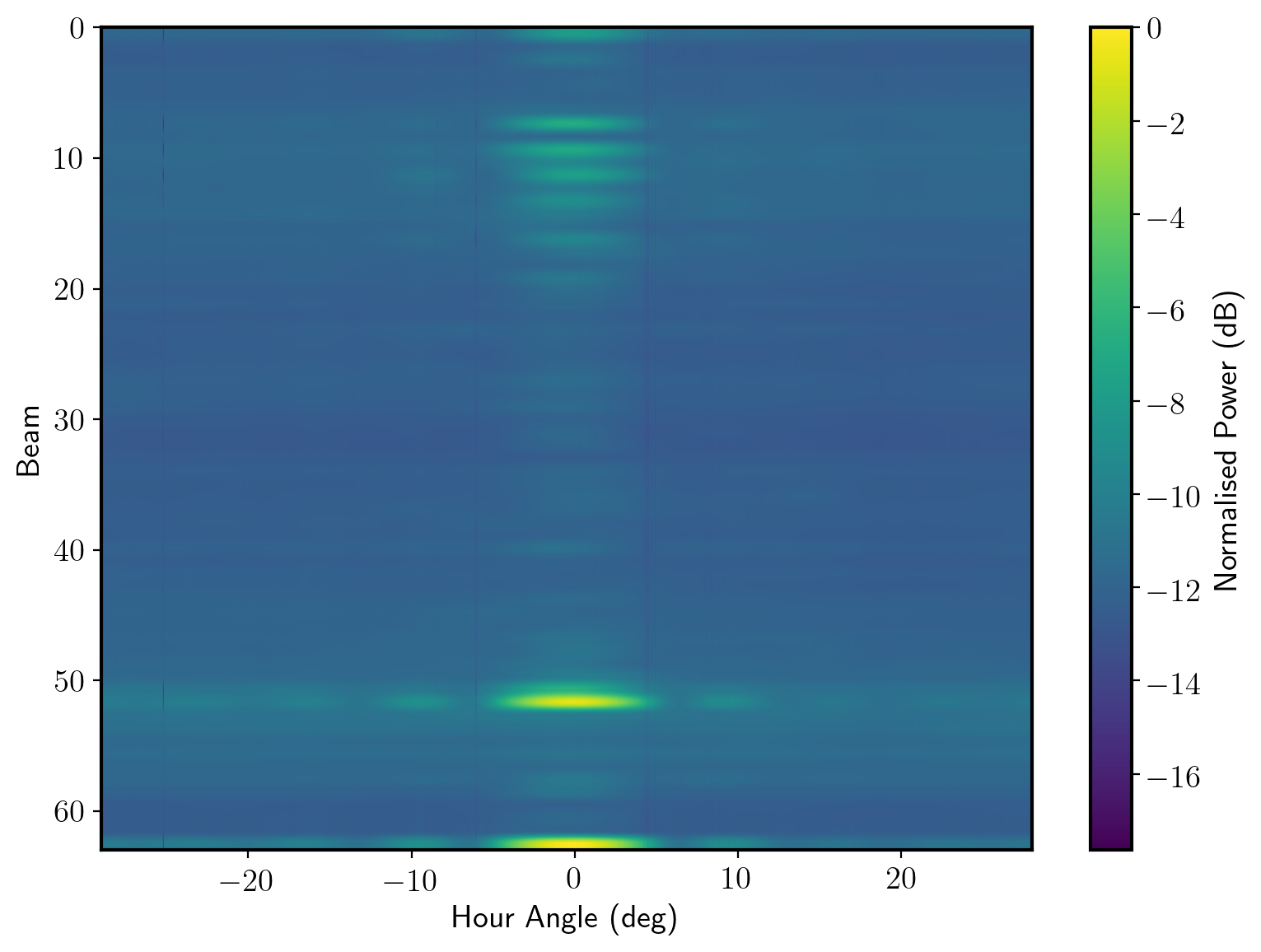}
      \caption{Cassiopeia A transit across all 64 synthesized beams. }
      \label{fig:casa_transit_all_beams}
  \end{subfigure}
  \begin{subfigure}[t]{0.45\textwidth}
    \includegraphics[width=\textwidth]{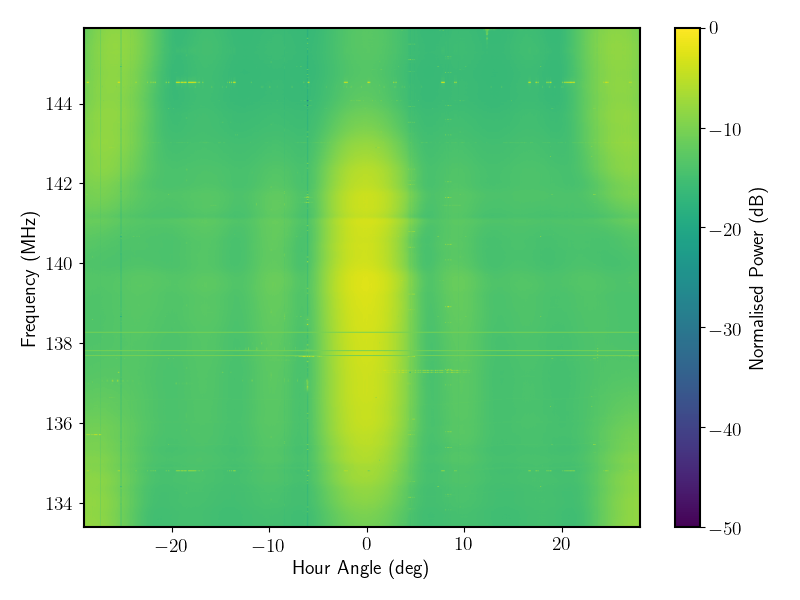}
      \caption{Waterfall plot of beam 64.}
      \label{fig:casa_watefall}
  \end{subfigure}
    \begin{subfigure}[t]{0.45\textwidth}
    \includegraphics[width=\textwidth]{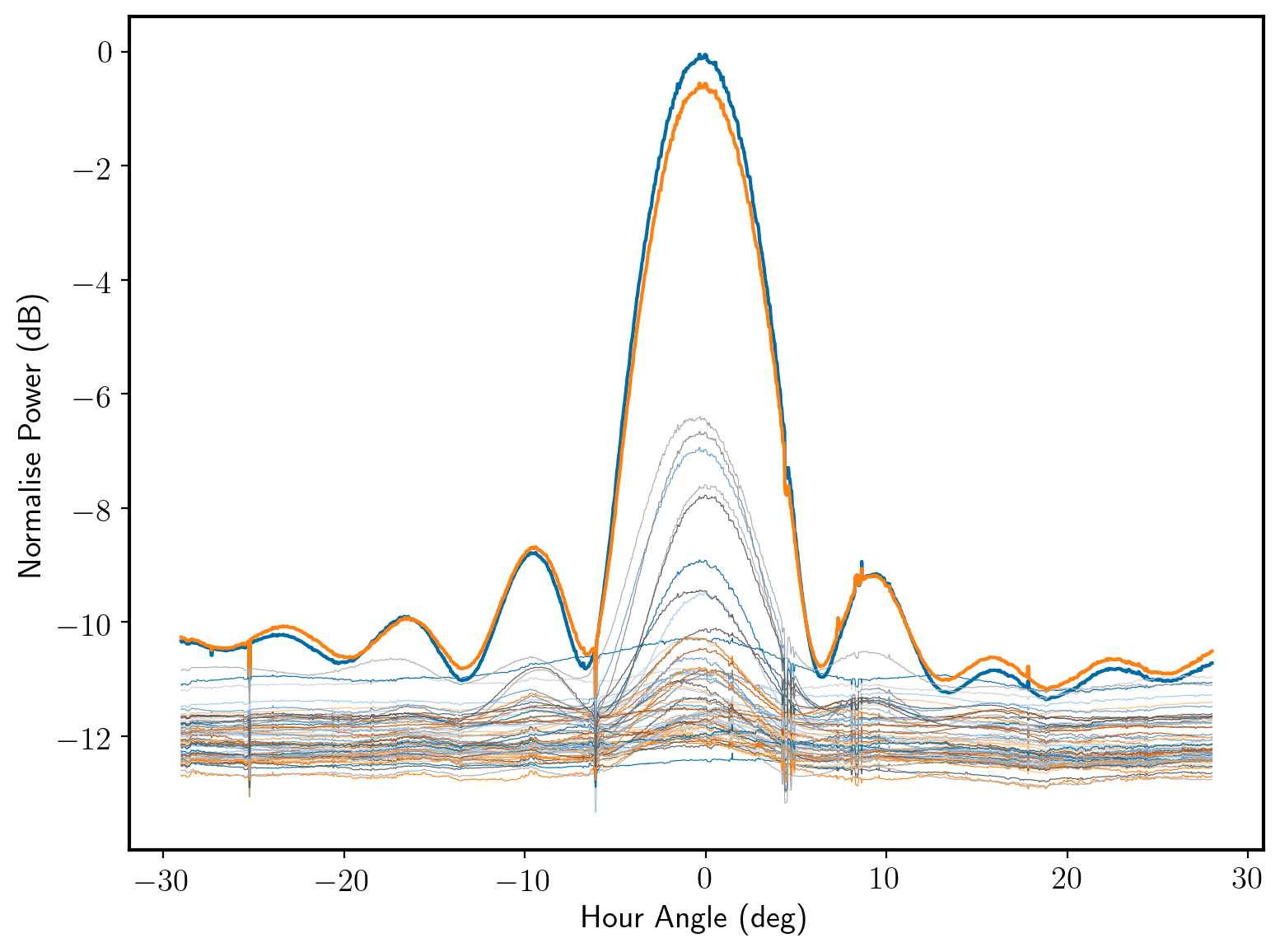}
      \caption{Cassiopeia A transit across all beams with central 2.5 MHz combined. Beams 64 and 53, having the highest SNRs, are highlighted.}
      \label{fig:casa_transit}
  \end{subfigure}
  \caption{One hour beamforming observation of Cassiopeia A, with the beamformer configured to generate 64 synthesized beam with an integration time of $\sim$1.34 s. See text for further details.}
  \label{fig:beam_observation}
\end{figure*}

The calibrated coefficients were then used in a 1 hour beamforming observation around the transit coordinates of Cassiopeia A performed on 14\textsuperscript{th} January 2021. The beamformer was configured to generate 64 North-South beams spread across 180$^{\circ}$ of altitude, as depicted in Figure \ref{fig:beam_pattern}, which shows the normalized far-field ampltiude for each of the beams. The integration time was set to $\sim$1.34 s. One of the beams, beam 64, was pointed at the highest altitude of the transit, such that it should have the maximum SNR. The beamformed data was saved to disk and a simple RFI filter was applied in post-processing. This filter first fits a polynomial across the band to generate a bandpass model and then compares this with the means of individual channels. The entire frequency channel is excised is excised if its mean is $\geq 5\sigma$. Finally, spurious RFI events are excised within each channel, and the bandpass is flattened.

Figure \ref{fig:casa_transit_all_beams} shows the power within each beam across time for the central channel. The beams with the highest power are beam 64, which pointed directly towards Cassiopeia A, and beam 53, which has an associated declination pointing closest to Cassiopeia A's (59.6$^{\circ}$). Figure \ref{fig:casa_watefall} shows the transit of the source in beam 63 for across the entire band, showing the behaviour of the instrument across frequency. For this and the subsequent plot, a median filter was applied to smooth out the beam. The sensitivity is greatest around the central $\sim$4 MHz, diminishing when moving away from this region. Additionally, reflections can be seen at the edges of the band, the cause for which is still under investigation. Finally, Figure \ref{fig:casa_transit} show Cassiopeia A transiting all the beam, outlining the beam shape for each pointing. In this case, the central 2.5 MHz (100 frequency channels) were combined. \citet{americo:mexart} provide a comparison of the observed array pattern with a modeled one. 

\begin{figure*}
  \includegraphics[width=\textwidth]{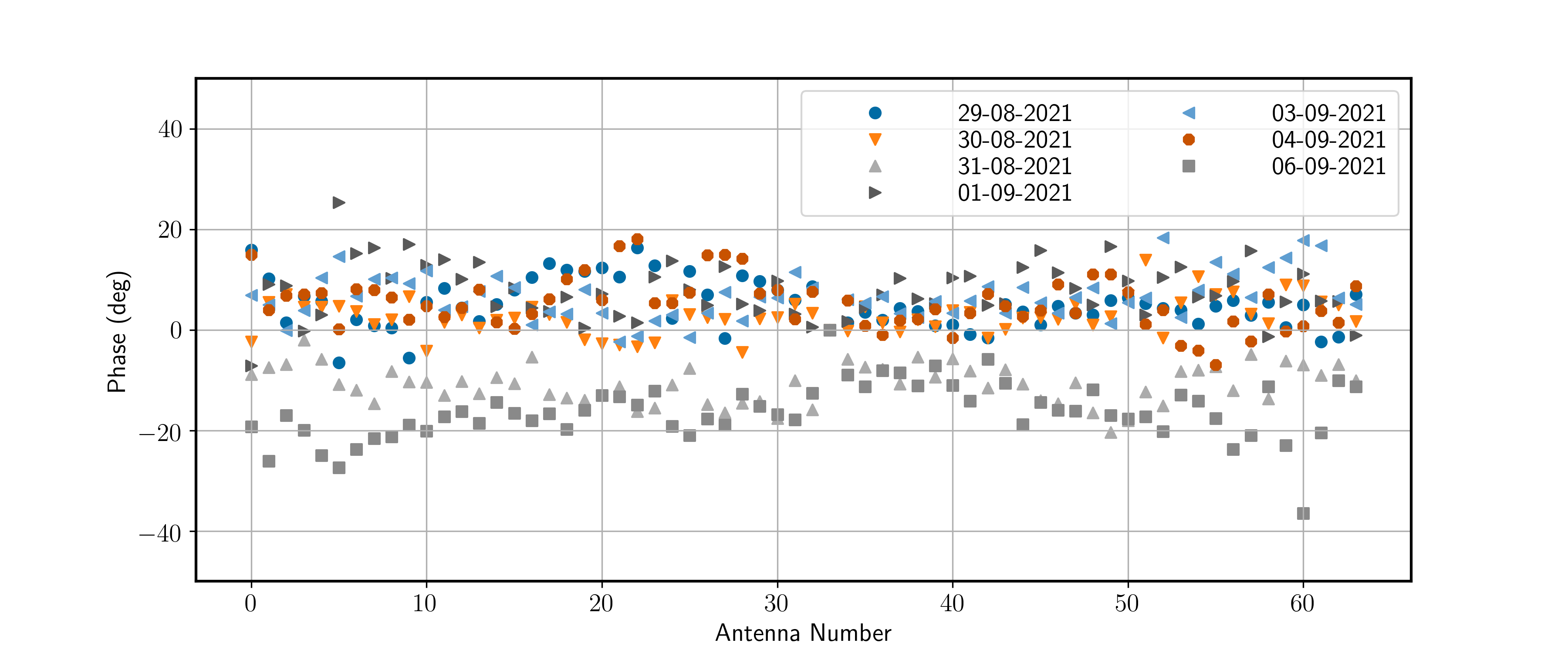}
    \caption{Per-antenna angle deviations, using antenna 32 as a reference, from 7 calibration observations using 3C123 performed over a 9 day period. The phase remains stable, varying by $\pm$ 20$^{\circ}$.}
    \label{fig:calibration_stability}
\end{figure*}

Once the performance of the system was verified, a stability test was performed. Daily calibration observations of the same astronomical source, 3C123,  were performed, generating the calibration coefficients for each. The stability of the system can be analyzed by studying how the calibration coefficients evolve in time. For each observation, the coefficients are phased relative to one of the antennas, in this case antenna 32, and the angle unwrapped along the antennas. Finally, each antenna's angle is normalised across observations, resulting in Figure \ref{fig:calibration_stability}. The observations here span 9 days, during which time the most of the deviation in the computed angle difference is $\pm$ 20$^{\circ}$. These deviations could be attributed to several factors. Apart from environmental conditions, including daily variations in prevailing atmospheric perturbations, the deviations observed can also be attributed to the analogue combination of rows of dipoles to form a single element. This thereby introduces an error to each element's real individual position, which is taken as a single point in Cartesian coordinates when calculating geometric coefficients.

\section{Conclusions}
\label{conclusion}

The main focus of this work has been on presenting the new digital backend for the MEXART radio telescope, showing the currently available features of the design, and preliminary analysis of system performance and stability. We have developed, deployed and tested a hybrid digital backend consisting of two FPGA-based digital boards performing digitization, downconversion and channelisation, and a GPU server responsible for beamforming, correlation, and monitoring and control. Additionally, we have developed the necessary software for general use, including observation management and scheduling, scripts and wrappers for the generated data products, and general monitoring and control. A web-based front-end was also developed to provide an easy-to-use and remotely accessible entry-point to the functionality of the system. This system is currently in regular use at the observatory, with daily IPS observations. We are currently investigating possible future upgrades to expand the use cases for the instrument.

\section*{Acknowledgements}

MEXART acknowledges partial support from CONACyT AEM Grant 2017-01-292684 and LN 315829. E. Aguilar-Rodriguez acknowledges support from DGAPA/PAPIIT project IN103821.


\bibliographystyle{mnras}
\bibliography{mexart_digital} 

\label{lastpage}
\end{document}